\documentclass[12pt,preprint]{aastex}

\usepackage{graphicx}

\begin{document}
\title{
An axisymmetric hydrodynamical model for the torus wind in AGN.
~II:  X-ray excited funnel flow}

\author{A. Dorodnitsyn\altaffilmark{1,2}, T. Kallman\altaffilmark{1}, and 
D. Proga\altaffilmark{3}}

\altaffiltext{1}{Laboratory for High Energy Astrophysics, NASA Goddard Space Flight Center, Code 662, Greenbelt, MD, 20771, USA}
\altaffiltext{2}{Space Research Institute, Profsoyuznaya st., 84/32, 117997, Moscow, Russia}
\altaffiltext{3}{Department of Physics and Astronomy, University of Nevada, Las Vegas, NV 89154, USA}
\begin{abstract}
We  have calculated a series of models of outflows from the obscuring torus in active galactic nuclei (AGN).
Our modeling assumes that the inner face of a rotationally supported torus is illuminated and heated 
by the intense X-rays from the inner accretion disk and black hole.
As a result of such heating a strong biconical outflow is observed in our simulations.
We calculate 3-dimensional hydrodynamical models, assuming axial symmetry, and including the 
effects of X-ray heating, ionization, and radiation pressure.  We discuss the behavior of a large family 
of these models, their velocity fields, mass fluxes and temperature,  as functions of the torus properties 
and X-ray flux.  Synthetic warm absorber spectra are calculated, assuming 
pure absorption, for sample models at various inclination angles and observing times.
We show that these models have mass fluxes and flow speeds which are comparable to those which have been 
inferred from observations of Seyfert 1 warm absorbers, and that they can produce rich absorption 
line spectra.
\end{abstract}
\section{Introduction}

One of the insights provided by observations of Seyfert galaxies and some quasars is the prevalence in their X-ray spectra of
spectral lines and bound-free continua from ions of intermediate-Z elements .
Early observations of Seyfert 1 galaxies using proportional counters and solid  state detectors revealed spectra 
with strong absorption features in the 0.1-10 keV range \citep{Halpern84}. 
These features were attributed mostly to the edges of hydrogen and helium - like oxygen. 
The term "warm absorber" was proposed owing to the fact that the observed X-ray absorbing gas has 
an electron temperature lower than it would be if a similar level of ionization were produced by collisional ionization.
However, more detailed spectroscopic studies were hampered by the limited X-ray resolution of the ASCA and ROSAT satellites.  
The grating spectrographs on the X-ray telescopes {\it Chandra} and {\it XMM-Newton} provide unprecedented 
spectral resolution up to $\sim10\, {\rm keV}$.
These show that X-ray spectra obtained from $\sim$ half of low-red-shift active galactic nuclei 
(AGN) contain many lines from ions of Fe, Si, S, O, Mg, and Ne, and that these are generally broadened and 
blueshifted by 100-500 km/s  \citep{Kaspi02, Steenbrugge05}.  
The presence of X-ray absorbing gas has been confirmed in the majority of AGNs which 
are bright enough to allow detections \citep{Reynolds97,McKernan07}. 
There is also a partial correspondence between UV and X-ray absorbers \citep{Crenshaw99}.

X-ray observations of warm absorbers are consistent with the 
Seyfert 1/Seyfert 2 dichotomy. For example,
the properties of the X-ray emission in the Seyfert 2 galaxy NGC 1068 
corresponds to the scattered emission expected from
warm absorbers in Seyfert 1 galaxies \citep{Kinkhabwala02}. 

Constraints on the position and dynamics of the X-ray absorbing gas can be deduced 
from the observed widths and virial arguments, 
and also from the variability studies of these spectra \citep{Behar03, Netzer03}.
These show an absence of correlated response of the warm absorber gas to rapid changes  
($\sim$ days) of the continuum.   This implies that the ionization time scale in the 
warm absorber gas is long ($\gtrsim$ months).
Combined together, the line blueshifts, widths, and time variability analysis favors 
an origin of the warm absorber gas at $R \gtrsim 1\, {\rm pc}$ away from the BH.
This estimate coincides with the likely location of absorbing matter responsible for obscuration in 
Seyfert 2 galaxies \citep{KrolikBegelman88}.
The existence of an outflow from the torus has been suggested by \cite{KrolikBegelman86, KrolikBegelman88}, 
and as the source of warm absorber flows by 
\cite{KrolikKriss1,KrolikKriss2}.

It is believed that this matter is in the form of  a molecular torus which is responsible for obscuring the 
broad line region in Seyfert 2 galaxies, and which is thought to exist  in most low and 
intermediate luminosity AGN \citep{AntonucciMiller86}.
A growing body of direct observational evidences advocates for the existence of the obscuring torus.
Mid-infrared  high spatial resolution studies of the
nucleus of NGC 1068 using the Very Large Telescope Interferometer have resolved a dusty structure 
which is 2.1 pc thick and 3.4 pc in diameter ~\citep{JaffeNATUR}. 
Observations support a multi-temperature model: the temperature of the warm component
was established to be 300 K and  inside of it a second,  compact and hot ($>$800K) component has been found. 
Further studies of NGC 1068 systematically reduced estimates of the temperatures of different components \citep{PonceletPerrin06}.
Observations of the Circinus galaxy, which is among the closest prototype Seyfert 2 
galaxies, also revealed a dense and warm $T\gtrsim 300$ K component at about 
$0.2\, {\rm pc}$ from BH and cooler $T<300$ K component at $1\,{\rm pc}$ \citep{Tristam07}.
If the hotter component is located closer to the X-ray source, it may be
attributed to the inner part of the torus, heated by the radiation of the
compact nucleus.
Although the evidence is strongest for nearby
active galaxies, there is also a strong motivation to think that within the same 
obscuring torus paradigm exist those quasars whose central regions are
heavily obscured by gas and dust (Type II quasars). Evidence for this 
comes from spectro-polarimetric observations by \cite{Zakamska06}.

This paper is part of a series whose main goal is to test the hypothesis that the torus
is the origin for the warm absorber flow.
Preliminary results of this work have been reported in 
~\cite{Dora08} (Paper 1), in which we presented the results from a sample model and showed 
that the adopted model is promising in explaining the warm absorber phenomenon.
In this paper we provide  more details of our methods, and display results of models 
which span the space of input parameters.  We present and discuss the 
hydrodynamic quantities which characterize our models:  mass fluxes, velocity fields, and 
temperature structure.  We also show sample X-ray spectra, which we will 
discuss extensively in a later paper of this series.

Our approach can be described as having three basic parts:
i) setting up initial conditions, which requires defining an initial torus configuration and making assumptions about the external source of radiation; ii) implementation of the wind driving force (local heating-cooling rates and radiation pressure force) and actual 2D hydrodynamical calculations. The latter includes the numerical solution of the time-dependent 2D (so called 2.5D) system of equations, which takes into account centrifugal forces, and radiation pressure and heating terms;
iii) calculating of the X-ray line spectra using a code which adopts Sobolev radiation transfer and ionization calculations for plasma in the intense 
X-ray field.
Each of these steps is described in what follows. 

\section{Governing equations}
We solve the following system of equations:

\begin{eqnarray}
&&\frac{\partial\rho}{\partial t}+\nabla \cdot (\rho {\bf v})=0\mbox{,}\label{eqCont}\\
&&\rho \left(\frac{\partial {\bf v}} {\partial t}+({\bf v}\cdot\nabla){\bf v}\right)
=-\nabla p-\rho\nabla\Phi+\rho\,{\bf g_{rad} } \mbox{,}\label{eqMoment}\\
&&\frac{\partial\epsilon}{\partial t}+{\bf \nabla}\cdot\left({\bf v}(\epsilon+p+\rho\Phi)\right) =
H\mbox{.}\label{eqEnergy}
\end{eqnarray}
These are the conservation equations for: mass, momentum and energy. Heating and cooling processes are 
described by the function $H({\rm erg\, cm^{-3}\, s^{-1} }) $; $\epsilon$ - is the sum of the kinetic and internal energy densities:
$\epsilon=\rho \, v^2/2+e$.
These equations should be supplemented by the equation of state which we assume to be polytropic: $P=K\,\rho^\gamma$, where $\gamma\equiv 1+1/n$, and  $n$ is the polytrope index and $P=(\gamma-1)\, e$. A one-component, one-temperature 
$T=P\mu/\rho\cal{R}$, where $\mu$ is the mean molecular weight per particle,
${\cal R}=8.31\cdot 10^7{\rm erg\, K^{-1} \, g^{-1}}$ is the universal gas constant and plasma with $\gamma=5/3$ is assumed to constitute the flow. All three components of the flow velocity ${\bf v}=(v_r, v_\theta, v_\phi) $ are calculated, assuming azimuthal 
($\partial/\partial \phi\equiv 0$) symmetry. 
Equations (\ref{eqCont})-(\ref{eqEnergy}) are cast in a non - dimensional form with the characteristic scales set by the properties of the plasma orbiting at 
a characteristic distance, $R_0$ from a black hole of mass $M_6$ ( in units of $10^6 M_\odot$). The characteristic scales are respectively:  $t_0=R_0^{3/2}/\sqrt{GM}\simeq 4.5\cdot10^{11}\,r_{\rm pc}^{3/2}\,M_6^{-1/2}\,({\rm s})$ for the time,  where $r_{\rm pc}$ is the distance in parsecs, and
$V_0=\sqrt{GM/R_0}\simeq 6.6\cdot 10^6 M_6^{1/2} \, r_{\rm pc}^{-1/2}\,({\rm cm\,s^{-1} })$ for the velocity.

\section{Forces driving the flow}
\subsubsection*{Heating and cooling of the gas}
The forces accelerating the wind in our model result from the gradient of gas pressure and from radiation pressure. 
The thermodynamic properties of X-ray heated gas depend on the spectrum of the incident radiation as
well as on the local atomic physics. 
Under the assumption of photo-ionization
equilibrium the thermodynamic state of photoionized gas
can be parameterized in terms of the ratio of radiation energy density to baryon density \citep{Tarter69}: 
 
\begin{equation}\label{smallxi}
\xi=4\,\pi\,F_{\rm x}/n\mbox{,}
\end{equation}
where $F_{\rm x}=L_{\rm x}e^{-\tau}/(4\pi r^2)$ is the local X-ray  flux, $L_{\rm x}$ is the X-ray luminosity of the nucleus, and ${\displaystyle \tau=\int_0^r\kappa\rho\,dr}$ - is the optical depth, and $n$ is the number density. We assume that the attenuation is dominated by Thomson scattering
 $\kappa=0.2(1+X_{\rm H})\simeq0.4\,{\rm cm^{2}\,g^{-1}}$, where $X_{\rm H}$ is the mass fraction of hydrogen,  and the factor $e^{-\tau}$,
accounts approximately for the attenuation of the radiation flux on the way from the source toward a fiducial point. 
The methods adopted in this paper for treating the effects of radiation are essentially the same as those 
described in \cite{Proga00,Proga07}, and 
have been applied to various problems in the study of AGN and X-ray binaries.

Assuming that the there is a fraction $f_{\rm x}$ of the total accretion luminosity $L_{\rm BH}$ available in X-rays  and that the disk radiates a fraction
$\Gamma$ of its Eddington luminosity $L_{edd}=1.25\cdot 10^{44}\,M_6$ 
we estimate:  $\xi\simeq 4\cdot 10^2 \cdot f_{\rm x}\,\Gamma\, M_6/ (N_{23}\,r_{\rm pc})$, where 
$N_{23}$ is the column density in $10^{23}$ ${\rm cm}^{-2}$. 
If the dynamical time within the flow is much larger than the characteristic time of the photoionization and recombination 
then the ionization balance is determined by the condition of photo-ionization equilibrium.
The rates of Compton and photo-ionization heating and Compton, radiative recombination, bremsstrahlung and line cooling are then
given by approximate formulas, modified from those of \cite{Blondin94}, for these processes:

\begin{equation}
\Gamma_{\rm IC}({\rm erg\,cm^{-3}\, s^{-1}})= 8.9\cdot 10^{-36}\,\xi\,(T_{\rm x}-4T)\mbox{,}
\end{equation}\label{heating1}
for the Compton heating - cooling;

\begin{equation}
\Gamma_{\rm x}({\rm erg\,cm^{-3}\, s^{-1}})= 1.5\cdot 10^{-21}\,\xi^{1/4}\,T^{-1/2}(T_{\rm x}-T)T_{\rm x}^{-1}\mbox{,}
\end{equation}\label{heating2}
for the photo-ionization heating-recombination cooling , and
for the bremsstrahlung and line cooling:

\begin{eqnarray}
\Lambda({\rm erg\,cm^{-3}\, s^{-1}})&=&3.3\cdot 10^{-27} T^{1/2}\nonumber\\
&+&(4.6\cdot 10^{-17} \exp(-1.3\cdot 10^5/T)\xi^{(-0.8-0.98\alpha)} T{-1/2}+ 10^{-24})\, \delta\mbox{.}\label{heating3}
\end{eqnarray}

\begin{figure*}
\includegraphics[width=500pt]{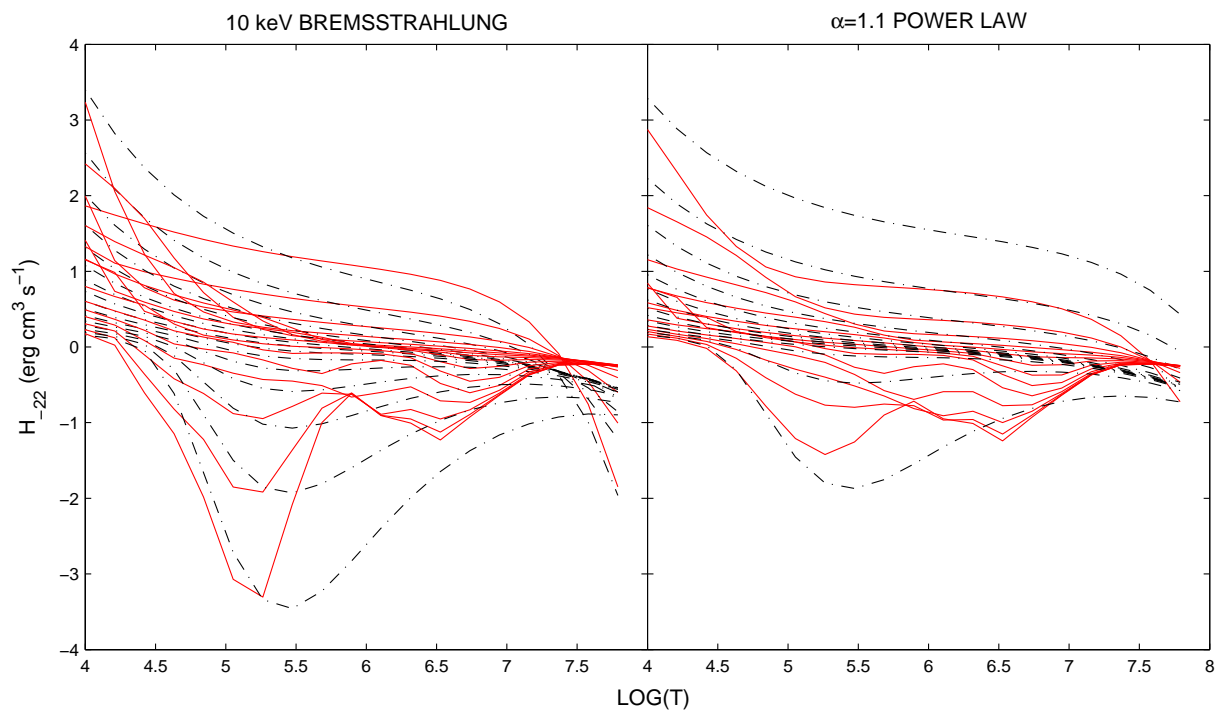}
\caption{
Comparison of the results from the XSTAR X-ray photo-ionization code and analytic 
approximations (\ref{heating1})-(\ref{heating3}). Curves are plotted for different values of $\xi$ ranging from $\xi=1$ (lower curves) to  $\xi=10^4$ (upper curves).
Vertical axis: total radiative  heating-cooling function $H=\Gamma_{\rm IC}+\Gamma_{\rm x}-\Lambda$ in units of $10^{-22}$
${\rm erg\, cm^3\, s^{-1}}$.
Horizontal axis: log(temperature).
Curves; solid: XSTAR; dot-dashed: analytic.
}\label{FigHeatCool}
\end{figure*}

These formulae have been originally derived for a 10 keV bremsstrahlung spectrum ($T_{\rm x}=2.6\cdot 10^7$ K) and were found to be in a reasonable ( $\sim 25\%$) agreement with numerical simulations \citep{Blondin94}.
Equations (\ref{heating1})-(\ref{heating3}) are slightly modified version of those of \cite{Blondin94}, which accommodates new atomic data.
Using the XSTAR code \citep{KallmanBautista01} we recalculated heating-cooling rates for the incident spectrum which is 
a power law with energy index $\alpha$, and found results essentially equivalent to those 
given by equations (\ref{heating1})-(\ref{heating3}).  
Notice that in the case of a bremsstrahlung spectrum a formal value of $\alpha=0$
should be used in (\ref{heating3}).
For a power law with energy index $\alpha=1.1$ the results 
differ by $\lesssim$ 30 $\%$ (see Figure \ref{FigHeatCool}).  Given these rates of energy deposition from the radiation to the flow, 
we write the total radiative heating-cooling function: $H=\Gamma_{\rm IC}+\Gamma_{\rm x}-\Lambda$. We have also performed several runs of our hydrodynamical models with different assumptions about heating-cooling, and found no important difference in the flow dynamics if using equations (\ref{heating1})-(\ref{heating3}) or the original formulae of \cite{Blondin94}, and also 
between bremsstrahlung and power law spectra for several values of $\alpha$. It appears that, for example, the effects of the optical depth are much more important.
That is, the difference between curves for the power law and the bremsstrahlung spectrum at small  $\xi$ (correspondingly high density) in Figure \ref{FigHeatCool} becomes unimportant.

\subsubsection*{The radiation pressure}

The radiation pressure force consists of the 
force due to continuum absorption 
$g_{\rm cont}=F_{\rm UV}\kappa/c$ and due to lines: 

\begin{equation}\label{radforcelines}
g_{\rm rad}=(F_{\rm UV}\kappa/c)\,M(t)\mbox{,}
\end{equation}
where $M(t)$ is the force multiplier \citep{CAK75}, and $ F_{\rm UV}$ is the local UV flux. We make use of the particular  form \citep{OwoCastRib}:

\begin{equation}
M(t)=k\,t^{-\alpha} 
\left( (1+\tau_{\rm max} )^{(1-\alpha)}
-1 \right)/\tau_{\rm max}^{1-\alpha}\mbox{,} \label{Mt1}
\end{equation}
where $t=\tau/\eta$ is the optical depth parameter,  $\eta=\kappa_l/\sigma_e$ is the line strength parameter, $\sigma_e$ is the Thomson cross-section,
and $\tau_{\rm max}=t\,\eta_{\rm max}$. 
A parameter $\eta_{\rm max}$ was introduced by \cite{OwoCastRib} and \citet{StevensKallman} in order to limit the effect of very 
strong lines.  That is, they assume a line number distribution which satisfies: 
$dN/(d\eta\,d\nu)\sim \eta ^{\alpha-2}\,\exp(-\eta/\eta_{\rm max})$, where $N(\eta,\nu)$ is the line number distribution. If $\eta_{\rm max}\to\infty$, so that lines are distributed as a power law, one recovers the result of \cite{CAK75}: $M(t)\sim k\,t^{-\alpha}$. In the opposite case of $\tau_{\rm max}\to 0$, the force multiplier is independent of $t$, and $M_{\rm max}\sim k\, \eta_{\rm max}^\alpha$. As a result of this maximum line strength cutoff a correction factor appears in the relation for $M(t)$, (\ref{Mt1}). 
The dependence of $k$ and $\eta_{\rm max }$ on $\xi$ has been numerically calculated and then fitted by the analytical formulae \citep{StevensKallman}:
\begin{equation}
k=0.03+0.385\exp(-1.4\,\xi^{0.6})\mbox{,}
\label{k}
\end{equation}

$$
\log_{10}\eta_{\max}=\left\{
\begin{array}{ll}
6.9 \exp( 0.16\, \xi^{0.4} )\mbox{,} & \log_{10}  \xi \le 0.5\mbox{,}\\
9.1 \exp(-7.96\cdot 10^{-3} \, \xi) \mbox{,} &\log_{10} \xi > 0.5\mbox{.} \label{eta_xi}
\end{array}
 \right.
$$
From these, one can see that  $M(t)$ can depend sensitively on the ionization parameter. Taking a fiducial 
$\alpha=0.5$ (the value, adopted in all our calculations), one finds that $M_{\rm max}=585$ at $\xi=0$, then has two local maxima: 
$M_{\rm max}=724$ at $\xi=0.3$, and $M_{\rm max}=743$ at $\xi=
3.1$.  $M_{\rm max}$ then drops to $1.7$  at $\xi=100$ and decreases gradually to $M_{\rm max}=0.01$ at $\xi=1000$.

\subsection{Initial configuration: rotating torus with arbitrary Compton optical depth.}
We begin from a rotating toroidal configuration which is in equilibrium 
in the external gravitational field of the BH.
The equation of state of the torus interior
is described by the polytrope $P=K\rho^{1+1/n}$.
The distribution of the density (or pressure) in the torus interior was given by \citet{PP84} (in what follows refer to 
PP-torus for short), who assumed that the distribution of the specific angular momentum inside the torus is constant. In our case such a torus would not be in equilibrium because of the radiation pressure from the central object. 
Thus we modify equilibrium equations of  \citet{PP84} to include the radiation pressure term. 
Since this cannot be done in a closed analytical form, we can write an approximate equation: 

\begin{equation}\label{PP-torus}
\frac{p}{\rho}\simeq\frac{1}{n+1}\left(\frac{1-\Gamma \,e^{-\tau(r)}}{r}-\frac{1}{2\,r^2\sin^2\theta}-C\right)\mbox{.}
\end{equation}
Note that equation (\ref{PP-torus}) must be understood as a bridging formula between two limiting cases: optically thin $e^{-\tau} \sim 1$ (in which case it is the PP-torus with $1-\Gamma$ reduced gravity), and optically thick, when $g_{\rm rad} \sim 0$ (PP-torus case). A constant $C$ in the equation (\ref{PP-torus}), parameterizes the distribution of the torus models and is connected with the distortion of the torus (this is described in more detail below). Including the radiation pressure reduces the effective gravity, and thus the torus gas needs less entropy to sustain it against vertical collapse.
In both of these limiting cases this equation is exact.

 Notice that the problem of toroidal equilibrium in the presence of heating (or other radiation transfer effects)
introduces a characteristic length scale through the optical depth $\tau$, 
leading to non-self-similarity of the model.
\noindent
Equation (\ref{PP-torus}) was derived by assuming that the distribution of the specific angular momentum inside the torus is constant. 
Choosing non-dimensional units and working in terms of $\varpi$, the cylindrical radius in units of $R_0$,
if we define the  non-dimensional density $\rho$ such that $\rho(\varpi=1)=1$, and the non-dimensional pressure $P$ and internal 
energy $e$ such that: $P=(\gamma-1)e$, and $e(\varpi=1)=e_{0}$ then
\newline ${
\displaystyle e_{\rm 0}=\frac{n}{n+1}\frac{(1-\Gamma)}{(0.5-C)^n}
\left(\frac{1}{x}-\frac{1}{2\varpi^2}-C \right)^{1+n}
}$. The inner and outer edges of the torus are located at $\varpi^-$ and $\varpi^+$, respectively.
Bounded configurations exist only for $0<C<0.5$ and the distortion of the torus is described by the parameter
$d=(\varpi^-+\varpi^+)/2=1/(2C)$. 
The boundary of the
torus  is matched to the exterior by the condition $P=10^{-6}$.
The PP-torus is unstable to {\it non-axisymmetric} perturbations \citep{PP84}. However, this effect cannot
be numerically investigated in the azimuthal symmetry which we adopt, since no signals can propagate in the $\phi$ direction. 
At $\varpi>1$, matter that constitutes the torus has an excess of angular momentum with respect to the local 'equilibrium' keplerian value,
$l(\varpi,z)>l(r)$, and vice versa, in the inner parts of the torus $\varpi<1$, 
$l(\varpi,z)<l(r)$. It is the internal pressure of the torus ~(equation \ref{PP-torus}) that
inhibits matter from settling to smaller (or larger, depending on angular momentum) orbits. 
The gas first evaporates from the part of the torus which is closer to the source of radiation
and  tends to settle at larger $\varpi$, as soon as the  back pressure  supporting it drops.

We begin our simulations from the stationary configuration determined from equation (\ref{PP-torus}). We follow the torus evolution as it is being heated by X-rays. No replenishing of the gas which constitutes the initial torus is provided:
Therefore, the torus will eventually lose all its mass and will completely evaporate.
However, in the regime we are looking for, the evaporation is not dramatic and does not significantly deplete the torus during the characteristic dynamical time.

In the following sections we will show that the existence and character of the 
flow from the heated torus depends critically on the geometry.  That is, it depends on the 
divergence of the flow streamlines, the strength and incident angle of the X-ray illumination, 
and on the direction of the effective gravity in the rotating frame of the torus.  The flow is 
intrinsically two-dimensional, and therefore cannot be adequately described {\it a priori} by 
1D models, such as those preformed by \citet{ChelNet05}.
Furthermore, the shape of the torus, and thus the launching surface for the flow, is affected
by the flow.  So the torus interior cannot be considered as a boundary condition 
(e.g. as in \cite{BalsKrolik}); we need to include it in the computational domain.

\section{Methods}
 For our computations we adopt a spherical-polar coordinate system $(r,\theta)$, extending the computational domain $\{r_i,\theta_j\}$  from $r_{\rm in}=0.01$
to $r_{\rm out}=50$ in radius, and from $0$ to $\pi$ in the polar domain making  no assumption about equatorial symmetry. 
The number of points in the radial, $N_r$,  and polar, $N_\theta$,  directions are taken to be equal: $N=140$, in low resolution and $N=300$ in high resolution grids. The $\{r_i\}$ grid is 
non-uniformly spaced, i.e.  $r_2=r_1+(r_{\rm out}-r_{\rm in})(k_r^{1/(N_r-1)}-1)/((k_r^{N_r/(N_r-1)}-1))$, and
$r_{i+1}=r_i+(r_i-r_{i-1})k_r^{1/(N_r-1)}$, for $i=2,N_r-1$, and the refinement factor is $k_r$=4.
In order to achieve better resolution of the flow itself rather than the torus interior we also adopt a polar grid which has non-uniform spacing $\delta\theta_i=\theta_i-\theta_{i-1}$, 
so that the maximum refinement is approached at
$\theta=\pi/4$: $\delta_{i+1}=\delta_{i}/k_\theta^{1/(N_\theta-1)}$ at $0<\theta<\pi/4$, and $\theta=\pi/4$: $\delta_{i+1}=\delta_{i} k_\theta^{1/(N_\theta-1)}$ 
at $\pi/4<\theta<\pi/2$ (and analogously spaced in the southern hemisphere). Boundary conditions are axially symmetric at  $\theta=0, \, \pi$ and 
outflowing at $r_{\rm in}$ and  $r_{\rm out}$

To solve numerically the system of hydrodynamical equations 
(\ref{eqCont})-(\ref{eqEnergy}) we use the code ZEUS2D ~\citep{StoneNorman92}. 
Note that the characteristic time of X-ray heating/cooling
can be much shorter than the dynamical time, which in such a case  introduces strong stiffness to the system of equations (\ref{eqCont})-(\ref{eqEnergy}). 
To overcome this difficulty, some modifications have been made to the code.
The most important one is the implementation of a fully implicit update of the energy in equation (\ref{eqEnergy}) just prior to the transport step in ZEUS2D.
Addionally, we account for the radiation pressure (equation (\ref{radforcelines})) term.
As an initial test we have evolved a toroidal distribution of matter for two rotational periods and found the configuration to be stable. 
The gas is illuminated by the incident X-ray radiation with a power law spectrum with an energy index, $\alpha=1$.
The heating/cooling rates are described by the  approximate analytical formulae give in equations (\ref{heating1})-(\ref{heating3}).

\subsubsection*{Warm absorbers}
We test output of our hydrodynamical models against the ability to predict warm absorber spectra.  
To do this we use the output from the hydrodynamical code, of $\rho,{\bf v}$, and $T$
as an input to the calculation of X-ray line and photoelectric absorption spectra.
The numerical code has been specifically developed for calculation of spectra in the X-ray domain and makes use of procedures 
developed for the XSTAR \citep{KallmanBautista01} code, while calculating the ionization structure and distribution of opacities, 
and treating the radiation transfer in the Sobolev approximation \citep{RybickiHummer83}.  
Although the goal of this paper is to show that pure hydrodynamic 2D models can produce warm absorber spectra,
we present here only sample  spectra, assuming pure absorption.  We postpone a more detailed discussion, 
including a full 3D transfer calculations, to a separate publication. 

\section{Results.}
The most important parameters which determine the 
properties of the warm absorber flow are the initial Compton optical depth $ \tau_\bot^{\rm C}=\tau( \theta=90^\circ )$ of the torus (or equivalently the maximum initial  torus density $n_{\rm max}$), and the distance from the BH,  $R_0$.
We also explore the dependence on $\Gamma$ and $d$.
Other parameters are chosen having some typical values:
the mass of the black hole: $M_{\rm BH}=10^6\,M_\odot$, 
the Compton temperature of the X-ray radiation $T_{\rm x}=10\,{\rm keV}$, and the fraction of X-rays and UV radiation 
$f_{\rm x}=f_{\rm UV}=0.5$. (For rotating flows exposed to a multi-temperature radiation, see e.g. \cite{Proga08}). 
The latter is consistent with typical energy distributions of the radiation close to the BH \citep{Laor97}.
We neglect any changes in the BH luminosity. The important thermal time scales within the flow, namely the Compton heating and cooling time, 
$t_{\rm x}$ and the dynamical time, $t_{\rm dyn}$, may be of the same order $t_{x}\sim t_{\rm dyn}\sim 10^{10}\,{\rm s}$.
This is discussed in more detail later in this section. Thus the outflowing gas may not be in thermal equilibrium and adiabatic losses are likely to be important. Notice that a nearly hydrostatic Compton heated corona can exist only at $r\lesssim R_{\rm IC}=GM_{\rm BH}\,\mu/{\cal R}T_{\rm x}\simeq 8\cdot 10^{16}(M_6/T_{\rm x,7})\,{\rm cm}$, where $T_{\rm x,7}$ is the Compton temperature in terms of $10^7$K. In all of our models the major flow is located at $r>>R_{\rm IC}$.

We have calculated 20 models, including combinations for: $\tau_\bot^{\rm C}=1.3 \,({\rm models\, A_i})$, and
40 (models $B_i$); $R_0=0.5,\,\, 1,\,\, 1.5$; $\Gamma=0.1,\,\, 0.3, \,\, 0.5$ all with $d=2.5$; and two models with
$d=5$ (models $C_i$).
 These are summarized in Table 1 where some of the characteristic results from the computed models are presented. 
In what follows we describe in detail the cases which best illustrate the most important results.  We also 
discuss the dependence of our results on parameters, based on the behavior of the ensemble of models.

The model $A_6$ 
is similar to that described in Paper 1, although the initial torus in the model which is described here has a different distribution of $\rho$ and $e$ (see equation (\ref{PP-torus}), and the discussion thereafter), and smaller 
$\tau_\bot^{\rm C}$. In Paper 1 this model has been described in detail. Calculations presented here reveal more details and confirm the conclusions of Paper 1.
We begin here by describing results from model $B_6$, and later discuss how it differs from model $A_6$.

\begin{figure}[htp]
\includegraphics[width=440pt]{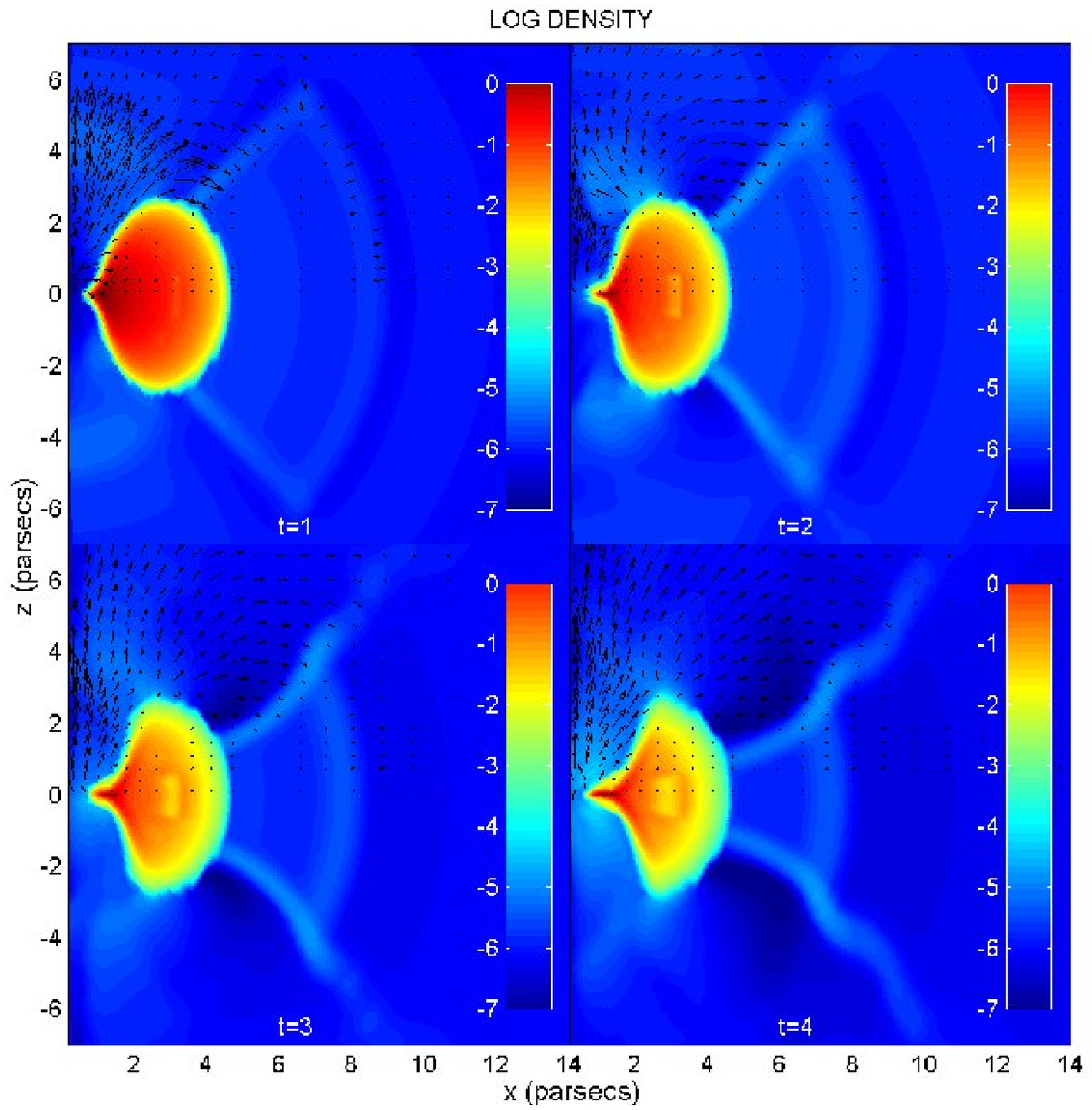}
\caption{Model $B_6$;
Color-intensity plots of the logarithm of the dimensionless density as a function of time which is given in orbital periods.  In the northern hemisphere this is superimposed with velocity vectors. 
Axes: distance in parsecs.
}
\label{Fig1_0}
\end{figure}
\begin{figure}[htp]
\includegraphics[width=340pt]{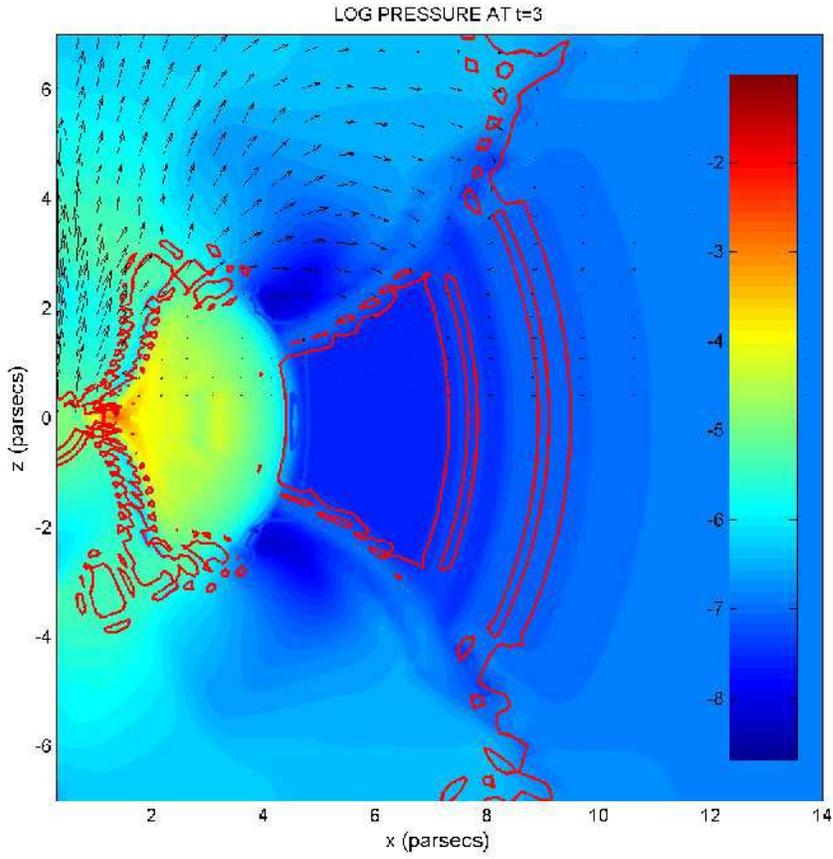}
\caption{Model $B_6$;
Color-intensity plots of the logarithm of the dimensionless pressure at $t=3$; at northern hemisphere superimposed with velocity vectors. 
The location of the sonic surface is marked in red.
Axes: distance in parsecs.
}
\label{Fig1_2}
\end{figure}
\begin{figure}
\includegraphics[width=340pt]{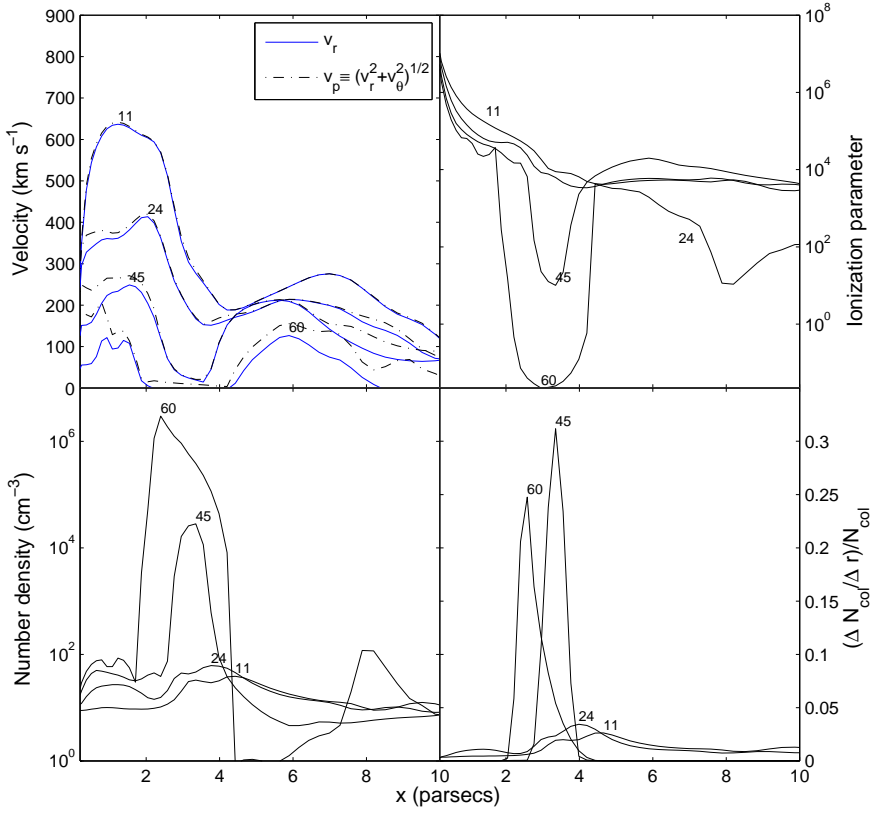}
\caption{Model $B_6$;
 distributions of  $v_r$, and the poloidal velocity  $v_p$ (upper left), 
$\xi$ (upper right),  $n$ (lower left), and the rate of growth of number density with radius (lower right) at time, $t=3$. Curves are marked by a inclination angle $\theta$.  
 Horizontal axis: distance in parsecs. 
} 
\label{Fig1_1}	
\end{figure}

Model, $B_6$ has $\tau_\bot^{\rm C}=40$, $R_0=1$ and $\Gamma=0.5$ and corresponds to a Compton thick ($\tau_\bot^{\rm C}\simeq 40$) torus having large $n_{\rm max}=10^{7}\,{\rm cm^{-3}}$ and mass $M_{\rm tor}=9.3\cdot 10^5\, M_\odot$.
Results are displayed in Figure \ref{Fig1_0}, where the evolution of the distribution of density is shown as a function of time (the density scale is such that 0 corresponds to $10^7\,{\rm cm^{-3}}$); Figure \ref{Fig1_2} where the distribution of pressure is shown at $t=3$ (the pressure scale is such that 0 corresponds to $4.7\cdot 10^{-4}\,{\rm dyn/cm^2}$) ; Figure \ref{Fig1_1} for various  quantities as a function of the inclination $\theta$;
Figure \ref{Fig2_0} (left panel), where the effect of the distortion parameter $d$ is demonstrated; and 
Figure \ref{Fig2_1} showing
horizontal 'slices' of  the velocity and temperature at constant height, $z$.
In the case of this model, the torus column is high enough to effectively screen the torus interior from penetrating X-rays.  This leads to 
formation of a nearly pure funnel flow, i.e. the torus interior, and hence the shape of the 
surface responsible for launching and collimating the flow, is essentially unaffected by X-ray heating on time scales $\lesssim t_{\rm rot}$.

Here and in what follows we discuss the time evolution of our models in terms of $t$, measured in units of the characteristic time of rotation, $t_0$.
After $t=1$  a high pressure region created by X-ray heating extends  to $r\simeq 4.5  -  5\,{\rm pc}$
throughout the area that is not shadowed by the high density torus. At this time the torus is located at $\theta<50^\circ$. 
The distortion parameter has a value $d\simeq 2.5$, i.e. the torus shape is almost unchanged from its initial value.  This is shown 
in the upper left panel of Figure \ref{Fig1_0}.
Within the part of the flow which is not shadowed by the torus, 
high temperature gas expands in a spherical bubble with radius, $r\lesssim 5.2$ pc in which the temperature is 
$T\sim 3 -  10\, T_{\rm vir}(r)$, where $T_{\rm vir}=2.6\cdot 10^5\, M_6/r_{\rm pc}$ is
the local virial temperature.
An axisymmetric region exists between $\varpi<0.75$ pc and $z<2$ pc where the temperature, $T\simeq 10\,T_{\rm vir}(r)$ .
That is,
high temperature, $T\sim 3\cdot 10^6\,\rm K$, but low density gas fills the torus funnel. 
The ionization parameter (equation \ref{smallxi})  in this region 
is $\xi\simeq 10^4 -  10^5$. 
The outer edge of the torus extends to $\sim 4.25 \,{\rm pc}$ in temperature, and to $\sim 4.5 \,{\rm pc}$ in density contours. 

Figure \ref{Fig1_0} (lower left) shows density and velocity fields for model $B_6$ at $t=3$.  
Figure \ref{Fig1_2} shows that a high pressure region 
expands to height $z\simeq 6 $ pc from the equatorial plane.  The torus  
inner edge is inferred from the temperature and density maps to be $\varpi^-\simeq 0.83\,{\rm pc}$.
Inside the of this radius, which we refer to as the torus throat, the temperature is
$T\simeq 10^6 -  10^7\,{\rm K}$.
A wide  nozzle 
with $(\varpi_{\rm max}-\varpi_{\rm min})/z_{\rm max}\simeq 2.12$, where
$z_{\rm max}\simeq 0.4\,{\rm pc}$
is formed, having inner radius of $\varpi\sim 0.85\,{\rm pc}$.  The torus outer edge is slightly shifted to 
$\varpi^{+}\simeq 4.5$pc.
The values of $\xi_{\rm min}$ (the minimum ionization parameter along a radial line) and the column density vary significantly with the inclination 
angle. Figure \ref{Fig1_1} shows the distribution of radial and poloidal velocity, $\xi$,  
density and the rate of growth of number density with radius  as function 
of $\theta$ at $t=3$ for model $B_6$.
Near the axis, $\xi_{\rm min}(\theta\simeq 20^\circ)=10^4$ and the column density is $N_{23}=10^{-3}$. 
Note that if $t_{\rm x}\gtrsim t_{\rm dyn}$, i.e. the gas is not in thermal equilibrium, then $\xi$ is not as meaningful as when $t_{\rm x}<< t_{\rm dyn}$.
When $t_x\gtrsim t_{\rm dyn}$ adiabatic losses strongly affect the temperature of the gas.
At  larger $\theta$,  the ionization parameter decreases:
$\xi_{\rm min}(\theta\simeq 25^\circ)=3\cdot 10^3$ and at higher inclination, $\xi$ gradually reduces from $\xi_{\rm min}(\theta=45^\circ)=12$,  eventually becoming $\xi_{\rm min}(\theta\simeq 60^\circ)=2.5$. 
At a critical angle, $\theta\sim 40^\circ$, a strong rise of the column density  reflects the fact that the line of site penetrates the dense torus body rather that through the wind (c.f. Figure \ref{Fig1_1}, lower right).
The  column density increases, from $N_{23}=0.3$ at $\theta=45^\circ$ to  $N_{23}\sim100$ at $\theta\simeq 60^\circ$, providing total obscuration.
Figure \ref{Fig1_2} also shows the position of the sonic surface determined by the 
relation  $v_p/c_s=1$, where $v_p=(v_r^2+v_\theta^2)^{1/2}$ is the poloidal velocity and $c_s=( {\cal R}\,T/\mu)^{1/2}$ is the speed of sound. Behind the torus a low entropy region exists which is bounded from the sides by a quasi-stationary shock.
The existence of this structure can be understood from the following considerations. If the flow were perfectly symmetric in both hemispheres, then it should have $v_z\equiv0$ at $z=0$, and the $z=0$ plane would be the equivalent of a rigid wall (reflecting boundary). Thus, if $v_z<0$ behind the torus the formation of a shock structure is anticipated.
Generally, this is the kind of picture one expects to observe from a supersonic wind flowing over a rigid obstacle.

At $t=5$ in model $B_6$ (Figure \ref{Fig1_0}, lower right),  the density maximum is located at $\varpi\simeq 2\,{\rm pc}$.
The inner edge of the torus does not shift significantly from the position it has at $t=3$:  $\varpi^-\simeq 0.75 \,{\rm pc}$ in density maps (and $\sim 1\,{\rm pc}$ in  temperature maps); the 
outer edge is at $\varpi^+\simeq 4.3\,{\rm pc}$. 
The temperature of  the torus interior is in the range $10^3 - 6\cdot 10^4$K.
A hot  flow is located near the axis, bounded from the sides by the torus throat, and having high temperature: $\sim{\rm few}\,\cdot  10^6$ K. 
A significant drop of ionization parameter $\xi$ from $\sim 6\cdot 10^3$ to
$\sim 6$, occurs again at  $\theta \gtrsim 45^\circ -  50^\circ$ (c.f. Figure \ref{Fig1_1}) , where the 
column density also rises from $N_{23}=0.04$, to $N_{23}=30$ at $\theta\gtrsim 60^\circ$. 
 The aspect ratio of the torus is: $\Delta=R_0/H\sim 1$ in accord with what is inferred from observations \citep{KrolikBegelman86, JaffeNATUR}.
At low inclinations, $\theta\lesssim 10^\circ$, everywhere in the wind the poloidal component of the velocity is determined by $v_r$.  However, at $\theta>50^\circ$
{\it inside} the torus throat, the  $v_\theta$ component is important,  i.e. $v_\theta\sim v_r$ at $\varpi<1\,{\rm pc}$.


Model $A_6$ has
$\tau_\bot^{\rm C}=1.3$, $R_0=1$, $\Gamma=0.5$ and is very similar to the model described in Paper 1. 
It differs from model $B_6$ in that the smaller optical depth of the torus 
interior cannot shield the gas from a extensive X-ray heating and the torus loses mass from large parts of its surface.
The initial maximum density of the torus is $n_{\rm max}=10^6\,{\rm cm^{-3}}$ corresponds to initial torus mass, $M_{\rm tor}=9\cdot 10^4\, M_\odot$.
Figure \ref{Fig2_1D} shows the distributions of  poloidal velocity, $\xi$,  
density and the rate of growth of number density with radius  as a function 
of $\theta$ at $t=3$ for model $A_6$ (in the same format as Figure \ref{Fig1_1}). During the evolution,
a region of high pressure extends from
$r\simeq 4.5  -  5 \, {\rm pc}$ at $t=1$ to  $r\simeq 12 \, {\rm pc}$ at $t=3$, and to
$r\simeq 20 \, {\rm pc}$ at $t=5$.
The inner edge of the nozzle shifts slightly from $\varpi^- \simeq 0.8 \, {\rm pc}$ at t=1 to
 $\varpi^- \simeq 0.83 \, {\rm pc}$ at t=3, and 
$\varpi^- \simeq 0.75 \, {\rm pc}$ at t=5. 
At later times the behavior of the model $A_6$ is similar 
to models $A_1$ and $A_5$, and can be inferred from Figure \ref{FigA1A5}

It has been mentioned that
in model $B_6$ much of the torus interior is opaque to penetrating X-rays.  Remarkably, the minimum nozzle cross-section doesn't change much at late times, implying that the mass-loss rate becomes quasi-saturated. 
Note that in the case of a 1D flow ${\dot M}$
is roughly set by the position of the sonic point, which in turn is set by gravity. In the case of a 2D nozzle, the mass-loss rate is determined by X-ray heating, gravity and the minimum nozzle cross-section. 
In the case of model $B_6$ the latter remains almost unchanged in time. We believe this model is probably most representative in showing the key features of X-ray excited flow.
However, models $A_i$ may generally have broader angular patterns in which a warm absorber spectrum is observed, as will be discussed below. Only comparing synthetic spectra with observations can answer the question of what model is more adequate in describing the phenomenon of warm absorbers.


\begin{tabular}{c  c c c c c c c c c c}
 Model &  $\tau_\bot^{\rm C} $  &  $R_0$  &  $\Gamma$ &  $d$ & 
$v_{\rm max, t=3}^{10^\circ}$  & $v_{\rm max, t=3}^{45^\circ}$ & $v_{\rm max, t=5}^{10^\circ}$ & $v_{\rm max, t=5}^{45^\circ}$ & 
$\rm \dot M_{t=3} $ & $ \rm \dot M_{t=5}$\\[5pt]
\hline
\hline
\tt  $\rm A_1$ & 1.3 & 0.5 & 0.1 & 2.5 & 516 & 155 & 624 & 332 & $4.09\cdot 10^{-4}$ & $6.54\cdot 10^{-3}$ \\

 \tt  $\rm A_2$   & 1.3 & 0.5 &  0.3 &  2.5  & 710 & 317 & 847 & 330 & $1.48\cdot 10^{-3}$ & $4.31\cdot 10^{-3}$  \\

 \tt  $\rm A_3$   & 1.3  & 0.5 & 0.5 &  2.5 & 707 & 267 & 760 & 291 & $2.34\cdot 10^{-3}$ &  $2.14\cdot 10^{-2}$\\

 \tt  $\rm A_4$   & 1.3  & 1 & 0.1 &  2.5  & 547 & 189 & 514 & 217 & $1.76\cdot 10^{-3}$ &  $1.66\cdot 10^{-2}$  \\

 \tt $\rm A_5$   & 1.3  & 1 & 0.3 &  2.5    & 526 & 179 & 605 & 343 &  $9.65\cdot 10^{-3}$ &  $6.34\cdot 10^{-2}$ \\

 \tt  $\rm A_6$   & 1.3  & 1 & 0.5 &  2.5 & 570
 & 235 & 670 & 337 & $2.02\cdot 10^{-2}$ & $1.8\cdot 10^{-2}$ \\

 \tt  $\rm A_7$   & 1.3  & 1.5 & 0.1 &  2.5& 360 & 197 & 413 & 230 & $6.21\cdot 10^{-3}$ &  $1.64\cdot 10^{-2}$\\

 \tt  $\rm A_8$   & 1.3  & 1.5 & 0.3 &  2.5  & 388 & 169 & 540 & 310 & $1.40\cdot 10^{-2}$ &  $5.68\cdot 10^{-2}$  \\

 \tt  $\rm A_9$   & 1.3  & 1.5 & 0.5 &  2.5  & 317 & 207 & 663 & 370 & $2.66\cdot 10^{-2}$ &  $1.23\cdot 10^{-1}$ \\

 \tt  $\rm B_1$ & 40 & 0.5  &  0.1  &  2.5  & 673 & 318 & 522 & 320 & $3.56\cdot 10^{-3}$ &  $7.38\cdot 10^{-3}$\\

 \tt  $\rm B_2$   & 40 & 0.5 &  0.3  &  2.5  & 590 & 257 & 1004 & 471 & $1.16\cdot 10^{-3}$ &  $1.87\cdot 10^{-2}$ \\

 \tt  $\rm B_3$   & 40 & 0.5 & 0.5 &  2.5  & 907 & 383 & 957 & 459 & $5.49\cdot 10^{-3}$ &  $2.56\cdot 10^{-2}$ \\

 \tt  $\rm B_4$   & 40 & 1 & 0.1 &  2.5  & 506 & 205 & 438 & 236 & $3.04\cdot 10^{-3}$ &  $1.53\cdot 10^{-2}$ \\

 \tt $\rm B_5$   &  40 & 1 & 0.3 &  2.5 & 536 & 216 & 587 & 276 & $8.71\cdot 10^{-3}$ &  $2.85\cdot 10^{-2}$ \\

 \tt  $\rm B_6$   &  40  & 1 & 0.5 &  2.5 & 641 & 271 & 676 & 324 & $1.55\cdot 10^{-2}$ &  $7.24\cdot 10^{-2}$ \\

 \tt  $\rm B_7$   &  40 & 1.5 & 0.1 & 2.5 & 395 & 179 & 496 & 187 & $1.39\cdot 10^{-2}$ &  $3\cdot 10^{-2}$ \\

 \tt  $\rm B_8$   & 40  & 1.5 & 0.3 & 2.5  & 541 & 185 & 610 & 329 &  $2.22\cdot 10^{-2}$ &  $7.17\cdot 10^{-2}$\\

 \tt  $\rm B_9$   & 40  & 1.5 & 0.5 & 2.5  & 547 & 248 & 602 & 347 & $3.2\cdot 10^{-2}$ &  $8.53 \cdot 10^{-2}$\\

\tt  $\rm C_1$   & 40  & 0.5 & 0.5 & 5 & 890 & 464 & 770 & 349 &$3.46\cdot 10^{-3}$& $1.16\cdot 10^{-2}$ \\

\tt  $\rm C_2$   & 40  & 1 & 0.5 & 5 & 789 & 788 & 772 &  770 & $1.35\cdot 10^{-2}$ & $8.01\cdot 10^{-3}$ \\

 \hline
\end{tabular}
\tablename{
1.  Models, for different initial $\tau_\bot^{\rm C} $,  $R_0$, $\Gamma$, and $d$ and results for
the maximum velocity, $v^{\rm\theta}_{\rm max ,T=time}({\rm km\,s^{-1})}$, where $\theta$ is the inclination angle; and the mass-loss rate, $\rm\dot M_{T=time}(M_\odot/yr)$. }

\subsubsection*{Mass loss within the funnel flow}

It is instructive to consider the distribution of variables within a horizontal cross-section
at a certain height above the equatorial plane. In so doing, we interpolate 
the solution from an $(r,\theta)$ - spherical grid to a  $(z,\varpi;\, 100\, {\rm x}\, 100)$ Cartesian grid.
Figure \ref{Fig2_1} shows the distribution of temperature and z-component of velocity, in terms of the escape velocity, $U_{\rm esc}=(2GM_{\rm BH}/r)^{1/2}$, at different heights for model $B_6$. 

A hot region extends to $\varpi\simeq 1$ pc at $z=0.2$, and to $\varpi\simeq 2$ pc at $z=1$.
The "funnel" can be seen in distributions of both temperature and velocity. At the X-ray heated boundary of this nozzle gas is being heated so that its temperature increases suddenly to $\sim 10^6 -  10^7$ K.
This fact reveals an analogy between the torus flow with X-ray excited winds in X-ray binaries ~\citep{Basko77,McCray75}; we discuss this further later in this section.
Notice that in our case the inner surface of the torus both serves as a copious source of a gas and as a collimating funnel.

\begin{figure}
\includegraphics[width=340pt]{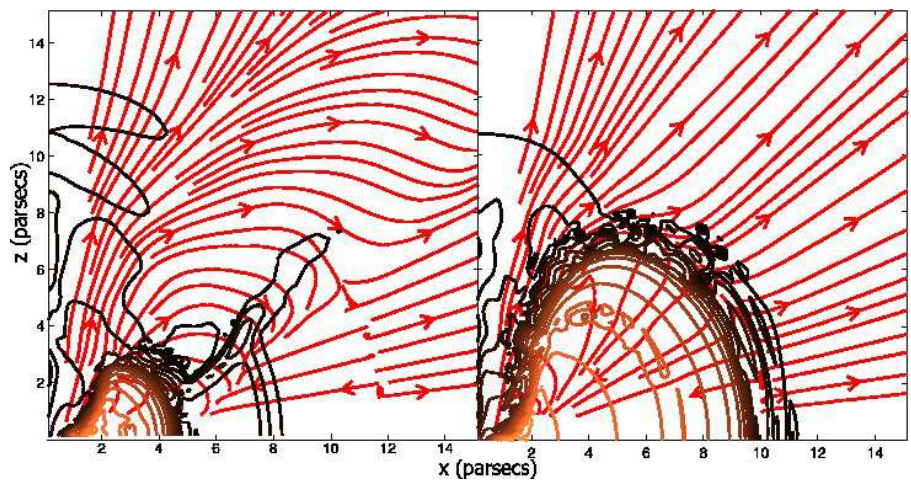}
\caption{The effect of distortion parameter $d$; Velocity streamlines superimposed on contours of the number density. Model $B_6$  (left panel),
model $C_2$ (right panel) at time, t=4. 
Axes: distance in parsecs. 
}
\label{Fig2_0}
\end{figure}

Figure \ref{Fig2_0} shows models $B_6$ and $C_2$ at $t=4$ and
Figure  \ref{FigA1A5} shows density and velocity streamlines for models $A_1$ and $A_5$ at $t=4$.
Notice, there is little difference between 
Figure \ref{FigA1A5} (left panel, Model $A_1$) and Figure \ref{FigA1A5} (right panel, Model $A_5$); the effect of smaller $R_0$ is partially compensated by the fact that $\Gamma$ is also smaller, thus 
reducing the effective gravity. If $\Gamma\simeq0$, then $\varpi^{-}\simeq 0.5$ (for $C=0.2$ in the equation (\ref{PP-torus})). However, when $\Gamma=0.5$, as in model $B_6$ (Figure \ref{Fig2_0}, left panel), the effective gravity at the innermost optically thin edge of the torus is reduced by half.
Figure \ref{Fig2_0} (right panel) shows a model with initially large distortion $d=5$ ($C=0.1$), model $C_2$ in Table 1.

In model $A_6$, 
a well-developed wind is observed in the vicinity of the high density torus, following the equal pressure contours; the maximum radial velocity is observed close to the axis at $v_{\rm max}(\theta\simeq 3^\circ)=700\,{\rm km}\,{\rm s}^{-1}$. As a general trend at $t=3$
the maximum velocity has a plateau at $20^\circ<\theta<50^\circ$, 
$v_{\rm max}=220\,{\rm km}\,{\rm s}^{-1}$, and lower values closer to the equatorial plane (Figure \ref{Fig2_1D}). The flow is approximately symmetric in both hemispheres.  At later times, $t=4$  and $t=5$, the behavior of the model is similar to $t=3$: namely, $v_{\rm max} (\theta\simeq 4^\circ, \,T=5)=
900\,{\rm km}\,{\rm s}^{-1} $, and on  the plateau being $v_{\rm max}\sim 380{\rm km\, s^{-1}}$.
The torus is losing mass in all directions, although with very different speed at different inclinations.
Because we are solving equations of ideal hydrodynamics (with only a small numerical viscosity), accretion through the inner boundary (at r=0) is negligible:  $\dot{M_{\rm in}}(M_\odot\,{\rm yr}^{-1}) < 10^{-8}$. 
The maximum mass flux  per unit solid angle 
$\dot{M}_\Omega^{\rm max}(M_\odot\,{\rm yr}^{-1}\, {\rm sterrad}^{-1})$ peaks at  $\theta\simeq 13^\circ$ at $t=3$, i.e. at much higher inclinations than $v_{\rm max}$, and at $\theta\simeq 55^\circ$,
$\dot{M}_\Omega^{\rm max}=0.01$ at t=1, $\dot{M}_\Omega^{\rm max}=2\cdot 10^{-3}$ at t=3, and $\dot{M}_\Omega^{\rm max}=0.02$ at $t=5$.
The total mass-loss rate at $t=3$ is $\dot{M}(M_\odot\,{\rm yr}^{-1}) \simeq 7\cdot 10^{-3}$.

The mass-loss rate is  $\dot{M}(M_\odot\,{\rm yr}^{-1})\simeq 2.4\cdot 10^{-2}$, at $t=4$, 
and  $\dot{M}(M_\odot\,{\rm yr}^{-1})\simeq 4\cdot 10^{-2}$, at t=5, and the change of the mass-loss rate with time is  $d\dot{M}/dt
((M_\odot\,{\rm yr}^{-1})/{\rm yr}) \simeq 10^{-6}$.
Comparing distributions of $v$ and $n$ we conclude, for example, that the apparent minima of $v_p\simeq v_r$ correlate (with a certain lag) with maxima of $n$ and vise a versa, reflecting conservation of mass flux. 

As in model $A_6$, the model $B_6$ funnel wind carries mass flux which doesn't change much during the evolution.
The maximum velocity is as high as $\sim 1000\, {\rm km s^{-1}}$ near the axis,
 and typically $200\lesssim v_{\rm max}\lesssim 600\,{\rm km}\,{\rm s}^{-1}$  at $15^\circ \lesssim \theta \lesssim 50^\circ$. 
The bulk of the gas, which potentially may produce warm absorber features, moves with comparable speed. However
the largest observed velocity in $B_6$ model is
$v_{\rm max}(\theta\simeq3^\circ)=1200\,{\rm km}\,{\rm s}^{-1}$, at t=5.
The mass-loss rate is $\dot{M}(M_\odot\,{\rm yr}^{-1}) \simeq 3.4\cdot 10^{-3}$, at t=3, and
 $\dot{M}(M_\odot\,{\rm yr}^{-1})\simeq 7\cdot 10^{-2}$, at t=5.

\subsubsection*{Spectra}
Computing absorption spectra is a key test for the warm absorber flow model. 
Several sample spectra are shown here, although the detailed discussion of methods and results of calculations of such spectra is postponed to a later paper.

Figure \ref{FigA6_spectr} shows the model $A_6$ spectrum observed at different inclinations.
This figure
shows the warm absorber spectrum at $t=3$ and at $t=4$.  At $t=4$ a rich X-ray line absorption spectrum exists in  the range $43^\circ \lesssim \theta \lesssim 52^\circ$, and in the range  $47^\circ \lesssim \theta \lesssim 55^\circ$ at later times, $t=5$. 

At $t=3$ the $B_6$ model predicts
a rich spectrum for $42^\circ \lesssim \theta \lesssim 47^\circ$. At later times a similar spectrum appears at  lower inclinations. Figure 
\ref{FigB6_spectr}
 shows the model $B_6$ spectrum observed at different inclinations
at $t=4$.
At $t=5$ the spectrum exists between  $45^\circ \lesssim \theta \lesssim 50^\circ$. Notice that the region of the funnel wind in this model is bounded by the area unshadowed by the torus:
$0^\circ \lesssim \theta \lesssim 40^\circ$.
At $\theta \gtrsim 30^\circ$ column density becomes $N_{23}\simeq 0.45$ and the ionization parameter is $\xi\lesssim 20$. At higher inclinations the X-ray flux in the $1<E<2$ keV range becomes severely absorbed.

Figure \ref{FigA3time_spectr} shows the evolution of the observed properties of the warm absorber flow with time (in the same time units) for model $A_3$. It can be seen that warm absorber spectra are changing slowly on a timescale $\Delta t\sim1$.
This is typical for most of our models and shows 
the range of times over which our solution can be considered as a representation of a 
steady state warm absorber flow.

A quantitative analysis  of our synthetic spectra and comparison with observations will be done in a later paper.  This is due in part to the 
need for full 3-dimensional treatment of the transfer and scattering of line photons, which we do not present here.  Rather, 
the spectra in figures \ref{FigA6_spectr}, \ref{FigB6_spectr} and \ref{FigA3time_spectr} are calculated assuming pure absorption.  We can 
calculate crudely some of the properties of individual lines, and show that these are generally consistent with observations.
A convenient way to do this is to discuss the profile of what is likely to be the strongest line in any synthetic spectrum, the $L\alpha$ line of OVIII.
In model $A_6$ {\rm at $t=3$} the full width at half-maximum (FWHM) of this line is  $\sim 200 \rm km\,s^{-1}$ . Closer to the BH, 
the maximum observed velocity is greater, i.e. models $A_3$ and $B_3$ give FWHM 
$\sim 400 \rm km\,s^{-1}$ at $\theta \sim 43^\circ$, and $40^\circ$, respectively. The centroid energy of the line is  at a blueshifted
velocity $(50-200)\rm km\,s^{-1}$ with respect to line center.  These velocities are less than those observed from, eg., NGC 3783, but
are comparable to those observed from other objects \citep{McKernan07}.  Such comparisons should also include the effects
of scattered emission, which may skew the line centroid and red edge, and which we have not considered here.

\subsubsection*{Analytical estimates of the mass-loss rate}
The mass-loss rate found from numerical calculations is in approximate agreement with theoretical expectations. The value of the mass-loss rate, ${\dot M}$, can be estimated by integrating the average mass flux $\langle j \rangle $ over the surface area of the torus exposed to X-ray radiation, $\Sigma\sim 2\pi^2\,R_0^2\,/ \Delta$, where $\Delta=R_0/H\sim 1$. $\langle j \rangle $ may be estimated using the same arguments as those of ~\citet{Basko77} and \citet{McCray75}. Namely, heating from a BH creates a narrow transition layer, a ''skin'' on the surface of the torus. There, temperature rises almost discontinuously from inner ''cold'' ($T\sim 10^4$, $T\lesssim T_{\rm vir}$), to outer ''hot'' ($T\gtrsim T_{\rm vir}$) value. This transition can be seen in Figure \ref{Fig2_1}.

Matching momentum, $p+\rho v^2$ and mass flux, $j=\rho v$ below and above this discontinuity, we obtain a well known relation: $j^2=(P_h-P_0)/(\rho_0^{-1}-\rho_h^{-1})$, where subscripts $0$ and $h$ refer to values below and above the discontinuity. Being heated, the gas expands and its specific volume, $V=1/\rho$ increases. Above the discontinuity the flow is assumed to be isothermal so that $P\sim 1/V$. In the P-V plane, the transition between points $P_0$, $V_0$ and $P_h$, $V_h$ goes through the straight line with an inclination,
$(P_h-P_0)/(V_h-V_0)>(dP/dT)_T$, and it follows $j^2<-(dP/dV)_T=\rho_h^2\,c^2_{s,h}$, where $c_s=( {\cal R}\,T/\mu)^{1/2}$ is the velocity of sound. Since $j=\rho_h v_h$, it follows $v_h<c_{s,h}$ and the flow immediately above the discontinuity is subsonic  \citep{Basko77,McCray75}.
 From the momentum conservation, $P_0\simeq P_m=
P_h+\rho_h v_h^2$, and the mass flux associated with such heating, can be estimated as
$\langle j \rangle =\displaystyle {\frac{P_m}{v_h\, {\cal M}_h(1+{\cal M}_h^2)}}$, where 
${\cal M}_h$ is the Mach number above the discontinuity,  $P_m$ is the pressure below the discontinuity, and 
for simplicity we assumed $v_h\simeq c_{s,h}$.

McCray \& Hatchett (1975) have
calculated the state of the gas in the optically thin layer of a stellar atmosphere heated by X-rays. 
From their results it follows that
the relation between $P_m$ and $F_{\rm x}$ can be cast in the form: $P_m=10^{-12}\alpha_{-12}\,F_{\rm x}$, where $\alpha_{-12}\sim 1$, reflecting 
the shape and effective temperature of the incident spectrum
 ~\citep{Basko77}.
Although it is essential (in order to obtain stationary transonic flow, correctly matching boundary conditions at infinity) that the flow above the discontinuity is subsonic, we assume that the sonic surface is located not far from the discontinuity, estimating $v_2=c_{s,h}$, ${\cal M}_h=1$. 
Next, we write: $F_{\rm x}={\displaystyle \frac{L_{\rm x}}{4\pi R^2}(1+A)}$, where $A$ is the effective X-ray albedo of the X-ray heated skin and we take $A=0.4$ (which we simplistically assumed to be optically thin), and assume $\mu=0.5$. Calculating $\dot M=<j>\Sigma$, we finally obtain:
\begin{equation}
 {\rm \dot M \,(\rm M_\odot/yr)}\simeq0.16\frac{f_{\rm x}\,\Gamma}{\sqrt{T_{h,6}}}\frac{M_6}{\Delta}\mbox{,}
\end{equation}
where $T_{h,6}$ is the temperature above the discontinuity in units of $10^6$K.
Inserting relevant parameters, such as $\Gamma=0.5$, $f_{\rm x}=0.5$,$R=1\,{\rm pc}$, $M_6=1$, $\Delta=1$,
and adopting the value of $T_h$ taken from our numerical model $A_6$, $T_h\simeq 10^6\,$K,
we estimate the mass-loss rate: $\dot{M}(M_\odot\,{\rm yr}^{-1}) \simeq 4\cdot 10^{-2}$. Comparing results from this approximate formulae with those summarized in the Table 1 we conclude that they are in good accord. Given the torus mass in the $A_6$ model $M_{\rm tor}=9.3\cdot 10^4\, M_\odot$, we conclude that it may sustain such mass loss for $\sim 1\cdot 10^6{\rm yr}$. The upper limit may be inferred from Table 1, and is found to be  $\sim 10^8{\rm yr}$.

\subsubsection*{Adiabatic loses}
\begin{figure*}
\includegraphics[width=340pt]{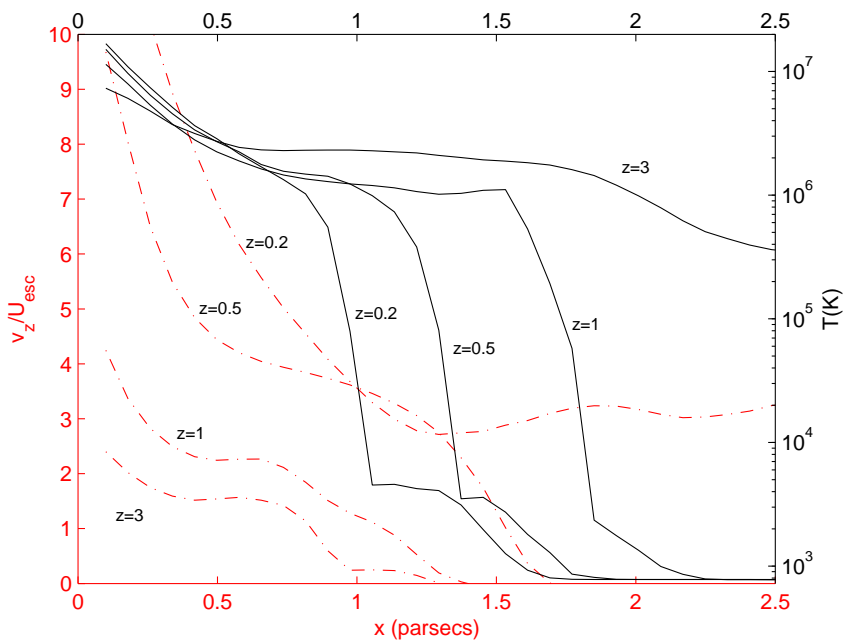}
\caption{Model $B_6$ at time, $t=3$.
Horizontal 'slices' of 
temperature and z-component of velocity are marked at each curve by the value $z$ in parsecs.
Vertical axis; left: z-component of the velocity in terms of local escape velocity;
right: temperature.
Horizontal axis: distance in parsecs.
Curves; dashed: velocity; solid: temperature. 
}\label{Fig2_1}
\end{figure*}

\begin{figure}
\includegraphics[width=340pt]{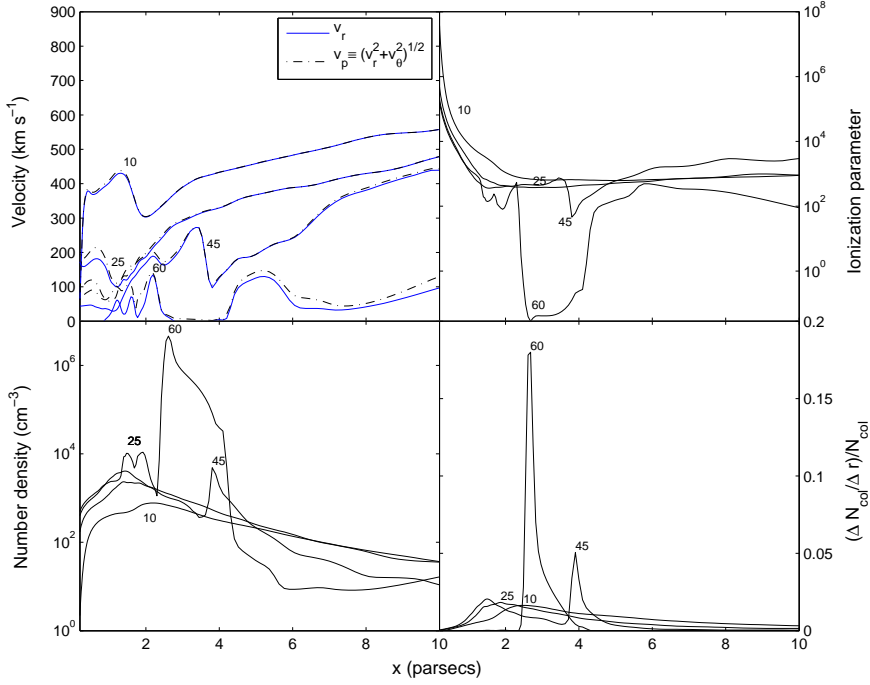}
\caption{Model $A_6$; As in Figure \ref{Fig1_1}
here are shown the 
distributions of  $v_r$, $v_p$ (upper left), 
$\xi$ (upper right),  $n$ (lower left), and the rate of growth of number density with radius (lower right) at $t=3$. Curves are market atop by an inclination angle $\theta$. Horizontal axes: distance in parsecs.
} 
\label{Fig2_1D}	
\end{figure}

\begin{figure}
\includegraphics[width=340pt]{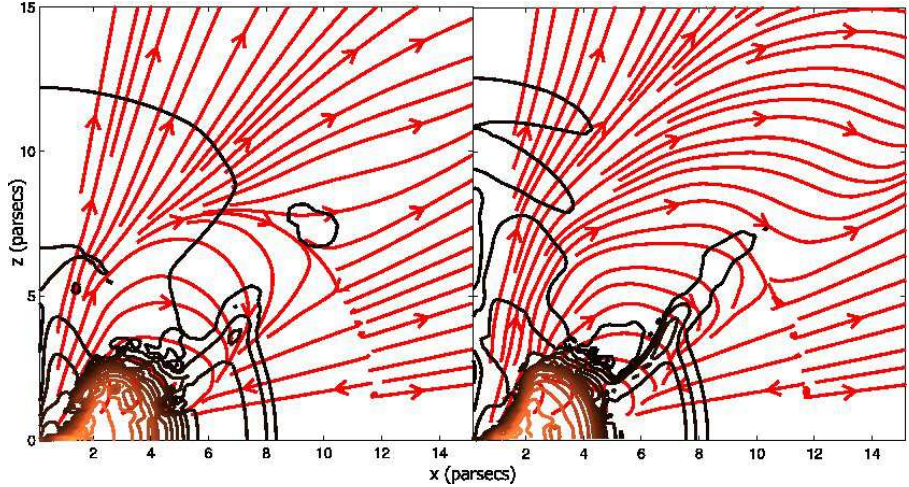}
\caption{The effect of $\Gamma$; velocity streamlines superimposed on contours of the number density. Model $A_1$  (left panel),
model $A_5$ (right panel) at time, t=4. Axes: distance in parsecs.
}
\label{FigA1A5}
\end{figure}

\begin{figure}
\includegraphics[width=340pt]{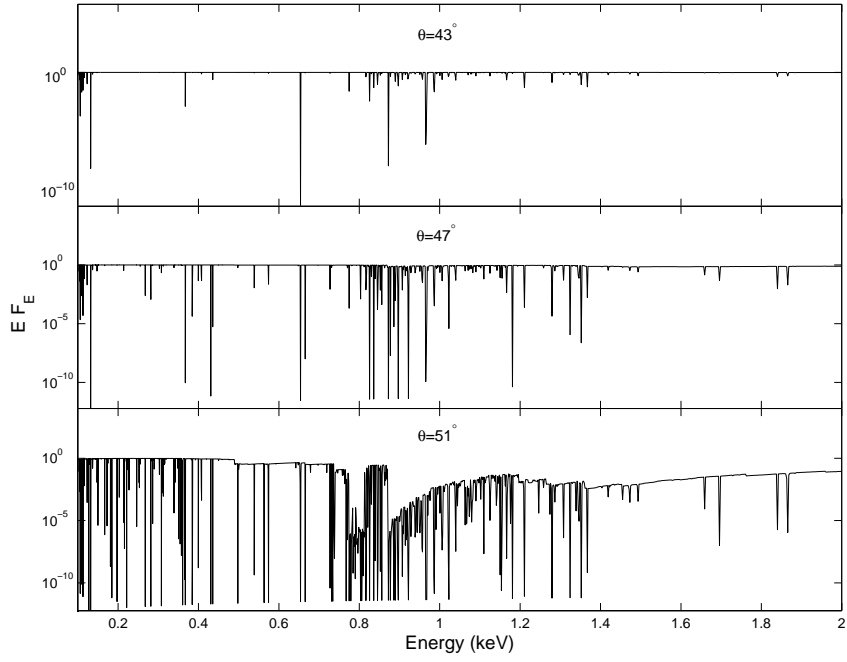}
\caption{Model $A_6$:
spectrum, observed at time, t=3.} 
\label{FigA6_spectr}
\end{figure}

\begin{figure}
\includegraphics[width=340pt]{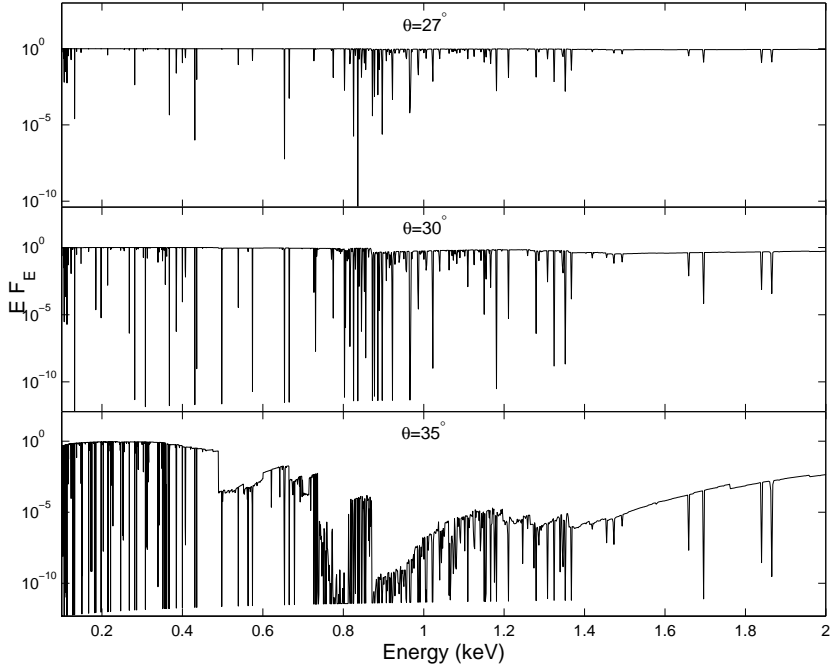}
\caption{Model $B_6$:
spectrum, observed at time, t=4.} 
\label{FigB6_spectr}
\end{figure}

\begin{figure}
\includegraphics[width=340pt]{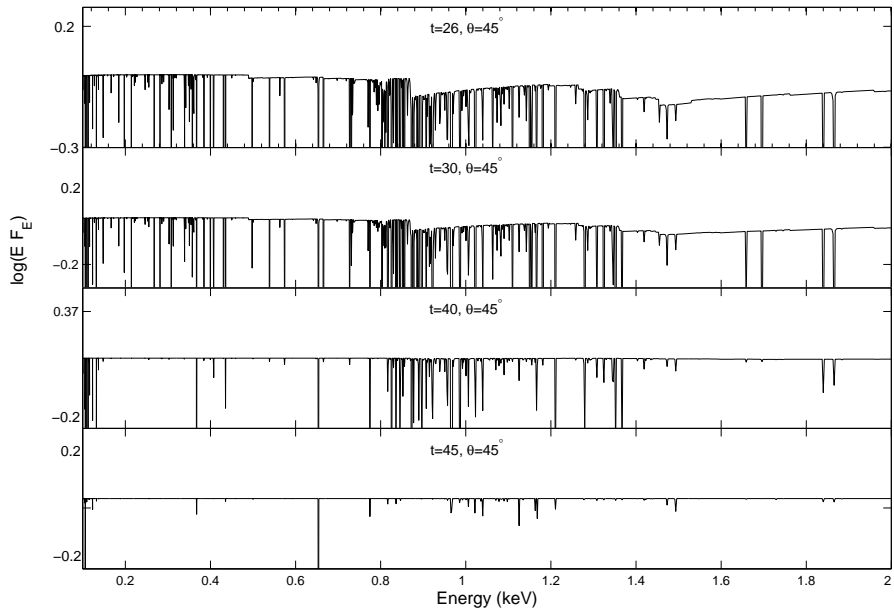}
\caption{Model $A_3$:
X-ray spectra, observed at $\theta=45^\circ$, as a function of time.
} 
\label{FigA3time_spectr}
\end{figure}

The characteristic time scale at which the energy is deposited to the flow via Compton processes, $t_{\rm x}$ can be cast in the form:
\begin{equation}\label{t_x}
 t_{\rm x}(\rm s)\simeq 9.4\cdot 10^{10}\,\frac{r_{\rm pc}^2}{\Gamma f_{\rm x}}
\frac{\tilde T}{\tilde{T}-1}\mbox{,}
\end{equation}
where ${\tilde T}\equiv T/T_{\rm x}$ and
$T_{\rm x}=2.9\cdot 10^7$K.
This should  be compared with the dynamical time, $t_{\rm dyn}$ of the flow:

\begin{equation}\label{t_dyn}
t_{\rm dyn}(\rm s)\simeq 4.3\cdot 10^{10}\, r_{\rm pc}\,\sqrt{\tilde T}\mbox{.}
\end{equation}
When $t_{\rm dyn}\lesssim t_{\rm x}$, the outflowing gas departs from thermal equilibrium and one must account for adiabatic losses, $\Lambda_{\rm ad}$, when calculating the 
temperature of the gas.
Notice  that the properties of the two-phase (or multiple-phase) gas are conventionally described by the S-curve on the $T-\Xi$ diagram \citep{Krolik81}, where $\Xi=F_{\rm x}/(nkTc)$ is the other form of the ionization parameter. That is on the $T-\Xi$ plot those places where $dT/d\Xi>0$ are stable to isobaric perturbations. Places where $dT/d\Xi<0$ are unstable. 
Including $\Lambda_{\rm ad}$, may significantly lower the temperature of the hot phase \citep{ChelNet05}. This temperature can be estimated by equating the Compton heating rate,
${\displaystyle \Lambda_{\rm IC}\simeq 4 k F_{\rm x}\frac{\sigma_e n}{m_e c^2}\, T_{\rm x}}$
 to the adiabatic losses rate, $\displaystyle \Lambda_{\rm ad}\sim \frac{v}{r}\rho c_s^2$. 
The flow near the funnel walls is less divergent than it would be in the case of a spherically-symmetric wind, in which case the latter expression is a factor of 2 larger. Assuming that above the discontinuity $v\sim c_s$, we obtain:
\begin{equation}
T_h(\rm K)\simeq 5.7\cdot 10^6
\left(\frac{f_{\rm x} \Gamma }{ r_{\rm pc}}\right)^{2/3}\mbox{,}
\end{equation}
which gives $T_h\sim 2\cdot 10^6$K, for parameters adopted  in this paper. This value is in good agreement with the value of $T_h$, which is found from $T(\varpi,z)$ distributions shown in Figure \ref{Fig2_1}. Three major regions within the funnel flow may be emphasized:  i) a ''discontinuity'' where temperature is rising
from the inner ''torus'' value to $T_h\sim 10^6$K; ii) a ''plateau'' where $T\sim T_h$ and thermodynamic characteristics of the flow result from the interplay between $ \Lambda_{\rm IC}$ and $\Lambda_{\rm ad}$; iii) region of hot, overionized flow where $T\to T_{\rm x}$. 

\subsubsection*{Returning current}
From Figures (\ref{Fig2_0}) and (\ref{FigA1A5}) we see that there exists a region, behind the dense torus, where outflow is switched to inflow. This gas rejoins the torus in the shadowed region.
For example,  taking the model $A_6$, and 
integrating the mass flux over the region where $v_p<0$, we obtain ${\dot M}_{\rm in}=4\cdot 10^{-6}\,M_\odot\,{\rm yr}^{-1}$ at t=3. That makes $\sim 6\%$ of the total accretion rate $\dot M_{\rm accr}$, required to maintain $0.5\,L_{\rm edd}$ luminosity of the BH, given the efficiency of accretion, $\eta=0.06$. At the same time, much more mass, $\sim 2\cdot 10^{-3}\, M_\odot\,{\rm yr}^{-1}$, is lost
within the funnel  ($\theta \lesssim 50^\circ$) in the X-ray excited wind.
Matter that is removed from the funnel is replaced by gas from the torus interior. Thus, a weak 
large scale convection flow is observed in the simulations. This effect is most clearly seen in models with large $d$, such as model $C_2$, shown in
Figure \ref{Fig2_0} and is due to a strong drop of $v_p$ as the outflowing gas is passing the shock wave front behind the torus (c.f. Figure \ref{Fig1_2}) and being unable to escape from the potential well.

\subsubsection*{Radiation force}
The dependence of the radiation pressure on the ionization parameter, $\xi$, is determined by equations (\ref{radforcelines}) and (\ref{k}). 
In the region of the fast flow the wind is too overionized for the radiation force to be important. This resembles Low Mass X-ray Binaries (LMXB) case in which the radiation pressure is also found to be insufficient to drive a significant outflow \citep{Proga02}.
The ionization parameter drops below $\sim 100$ at $\theta\gtrsim 40^\circ$ - the value determined by the torus aspect ratio, $\Delta$. Thus the radiation pressure may be of importance at higher $\theta$ and at these inclinations its relative strength is determined by the attenuation of the X-ray and UV fluxes.
For model $A_6$ we have $\tau\simeq 6$ at $\theta\simeq 90^\circ$ and 
$\tau\simeq 1$ at $\theta\simeq 60^\circ$; i.e. the torus becomes Compton thin at 
$\theta\lesssim 60^\circ$. The radiation pressure exhibits complicated behavior with varying $\theta$, having multiple maxima and minima. The force multiplier, $M(t,\xi(\theta))$ peaks at $44^\circ$ at $r\simeq 2$, where $g_{\rm rad}/g_{\rm grav}\sim 5$. Generally, two maxima of $g_{\rm rad}$ are observed at a given $\theta$ along a radial line. The second peak becomes smaller at higher inclinations, i.e. in models $A_i$ the radiation pressure is determined mainly by the properties of the X-ray heating (i.e. $\xi(\theta)$) rather than by the attenuation of the UV flux.
At higher $\theta$ smaller maxima occur at smaller $r$; the inner skin of the torus exerts considerable radiation pressure, although at large $\theta$ it is opposed by the back pressure of the torus interior. We calculated a model which has the same parameter values as model $A_6$ but with $g_{\rm rad}\equiv 0$. At $t=3$ this model gives $v^{\rm max}(\theta=10^\circ)=564\,{\rm km\,s^{-1}}$ and $v^{\rm max}(\theta=45^\circ)=194\,{\rm km\,s^{-1}}$. Comparing with Table 1 values we see that for the range of angles where warm absorber flow is observed the radiation pressure doesn't play a major role in the flow acceleration.
\newline
\noindent
In models $B_i$ the attenuation is much stronger than in models $A_i$ and consequently the secondary maxima of $g_{\rm rad}$ which were observed in models $A_i$ are suppressed by the $e^{-\tau}$ attenuation. The radiation pressure is important only on the skin of the torus but almost everywhere points in the wrong direction, opposing the back pressure of the torus interior. Only at $\theta\sim 45^\circ$ it points in the direction tangential with the torus surface, but as $r\sim 3$pc the density drops and $\xi$ rises so that $M(t,\xi)$ becomes small.

\subsubsection*{Dependence on $\Gamma$, $R_{c,0}$,$\tau_\bot^{\rm C}$, and $d$}
If the interior of the torus is optically thick to X-rays then the torus loses mass mostly from the surface, much as in the 'self-excited wind' scenario for X-ray binaries
\citep{Basko77}. As shown above, in such a case the torus throat serves as a funnel 
and the gas is injected to the flow from the funnel walls.

Notice that the location of the narrowest part of this funnel determines the characteristic terminal speed of the wind. 
In order to explore this, we have made a set of runs similar to models $A_3$ and $A_6$, but with reduced $\tau_\bot^{\rm C}$. For model ${\tilde A_3}$ which has $\tau_\bot^{\rm C}=2$ and $R_0=0.5$, we find that for $\theta=10^\circ$ and $t=4.5$, the maximum velocity $v_p^{\rm max}$ equals $738\, {\rm km\, s^{-1}}$. For model ${\tilde A_6}$ which has $\tau_\bot^{\rm C}=2$ and $R_0=1$, we find $v_p^{\rm max}=432 \,{\rm km\, s^{-1}}$ for the same $\theta$.
If in the latter model we make optical depth smaller, $\tau_\bot^{\rm C}=1$ we 
obtain: $v_p^{\rm max}(T=4.5,\,\theta=10^\circ)=400 \,{\rm km\, s^{-1}}$.
This shows, in accord with our expectations of the mass flux conservation, that the torus is losing mass from deeper inside. 
As shown in Table 1, reducing $R_0$ has the effect of increasing the maximum velocity.  An increase of $\Gamma$ has the same effect. However, this maximum velocity may be observed at a different inclination. Increasing the distortion parameter, $d$ has an effect of some increase of the maximum velocity, redistributing $v_p^{\rm max}(\theta)$ to higher inclinations.
From numerical solution we notice that the torus aspect ratio, $\Delta=r/H\sim1$ does not strongly influence the  evolution. That is because it is the most inner part of the throat which  determines the dynamics of an evaporative flow. This inner throat is located at 
high $\theta$ so that it remains optically thick most of the time.
Numerical experiments confirm that the geometry of this inner throat remains approximately unchanged in time. 

Figure \ref{Fig1_0} shows that the geometry of the innermost part of the torus, i.e. the densest part (roughly located between 0.5 and 2 pc) shrinks considerably in the vertical direction during the process of the evolution. This is the result of the joint action of the radiation pressure and the back pressure of the hot evaporative flow. 
This is particularly interesting as it resembles the geometrically thick outskirts of  AGN accretion disks which are known to be unstable to self-gravity
\citep{KolykhalovSyunyaev80,ShlosmanBegelman89}. 
The physics of such systems is complicated, and is subject to various possible competing effects.
The self-gravitating instability may operate also in the torus body, perhaps leading to a dynamical system of molecular-dusty 
self-gravitating clouds (as in \citet{KrolikBegelman88}). If this is the case, the optical depth of the torus, $\tau_\bot^{\rm C} $ is crucial as in the optically thin case the torus will effectively cool and collapse to a thin disk with subsequent star formation \citep{Toomre64}. In the other extreme ($\tau_\bot^{\rm C} >>1$) the released energy can go to increase the velocity dispersion of the clouds, effectively supporting the torus thickness ~\citep{Paczynski78}. 
Strong IR radiation pressure exerted on these clouds, which can come from internal reprocessing of X-rays, can produce significant 
vertical force \citep{Thompson05,Honig07}, and may suppress the self-gravity instability and at the same time provide  pressure support against vertical collapse \citep{Krolik07}.
Vertical support, and partial suppression of gravitational collapse, may also be provided by radiation pressure from star formation within the torus or the obscuring flow
\citep{Wada02}.
Further heating and loosing mass induces a torus to expand and change of shape.

We have calculated models $B_6$ and $A_6$ with 100x100 resolution further in time to learn the late time evolution.
At $T=17$ in the $B_6$ model the torus has two extended lobes in both hemispheres with an opening angle ${45^\circ}$. They have a certain degree of asymmetry with respect to the equatorial plane. The shape of the obscuring structure no longer resembles the initial torus; the column densities are in the range of $N_{23}=10$ at $30^\circ$, and $N_{23}=10^{3}$ at $30^\circ$. The radial velocities in this structure are in the range of $200-400 {\,\rm km \, s^{-1}}$.
Our model does not allow for the replenishing of the torus; obviously the torus will evaporate completely if given enough time.
Thus, in the the model $A_6$ the torus evaporates completely by the time $T=15$.
These results imply that in order to get a quasi-stationary warm absorber flow  the replenishing time should be of the order of the mass-loss time. The whole torus configuration may be unstable in a secular sense; the instability is driven by the long characteristic time of the {\it global} torus heating/cooling (due to expansion, winds, radiation loses), advection of heat in the torus body by internal flows etc. For example, the mass-loss rate, ${\dot M}\sim \Sigma$; the surface area $\Sigma$ increases during the torus expansion. If after some time of extensive heating, the torus separates into several parts, further mass loss will increase due to the larger total surface area of the fragments.

\section{Conclusions}

We have studied X-ray excited winds from the putative gas-dusty torus in AGN. 
We approach this problem using numerical methods combining detailed hydrodynamical modeling with calculation of the warm absorber spectra.
Hydrodynamical calculations include two-dimensional, 
axially-symmetric rotating flow, driven primarily by X-ray heating.
Compton, bremsstrahlung, and photoionization heating/cooling processes were taken into account as well as the radiation pressure force, which was calculated in the Sobolev approximation. 
A code combining XSTAR for photoionization calculations with 
the Sobolev radiation transfer has been developed for the calculation of the spectra. 

We find that a rotationally supported torus heated by radiation from the inner accretion disk and  black hole can indeed be 
a source of the material we observe in the warm absorber flow.
We find that the inner throat of the torus is not only important as a source of the gas but also because it creates a funnel for the outflowing wind. 
This leads generally to larger velocities within the funnel, and different velocity distribution within the warm absorber flow from those 
derived from models based on spherically-symmetric  winds.
The wind mass-loss rate within the funnel is not very sensitive to the details of the initial torus distribution and approaches $\sim 0.02-0.09\,
{\dot M}_\odot \, {\rm yr}^{-1} $.  
Strong X-rays heat the gas within the funnel, producing a fast, $\sim 1000\, {\rm km}\, s^{-1}$, ionized flow near the axis, and slower, $\lesssim 500\, {\rm km}\, s^{-1}$, flow closer to the funnel walls. This is where optical depth effects become important and a warm absorber spectrum is produced. Using methods developed in studies of X-ray binaries we were able to estimate the mass-loss rate from such funnel flow, finding it to be in a good agreement with our numerical solution.

The funnel flow is found to be promising with respect to obtaining high velocity warm absorber flows. 
What is beyond the scope of our models is the possibility of having multiple phases in such high velocity flow, 
on spatial scales smaller than our grid resolution.  Our treatment of the gas thermal properties will produce 
two-phase behavior at our grid resolution; we do not find this behavior, owing to the fact that 
the cooling timescales are generally too long.  The answer to the question of whether 
there can be high velocity 'bullets' or  'embedded clouds ' on length scales smaller 
than the resolution of the
grid is related to the problem of the origin of broad and narrow  UV/optical line emitting clouds, 
and requires different computational methods from those employed here. 

Our models which have initial Compton depths $\tau_\bot^{\rm C}\gtrsim 1$, aspect ratio $R_0/H\sim 1$, and located at $0.5\lesssim r \lesssim 1.5$ pc predict warm absorber spectra, thus confirming the main conclusion made in paper 1. The existence of such spectra depends on the fact that
the flow is intrinsically two-dimensional, meaning that 
both the dynamics of the funnel flow is different from 1D models 
and optical depth effects are important as they strongly depend on inclination. The latter point requires that we include the entire torus in the computational domain rather than considering it as a boundary condition.
The distribution of the ionization parameter, $\xi$ depends strongly on $\theta$, further confining the range of angles where conditions are right for the warm absorber flow to be observed.
In most of our models warm-absorber-like spectra are produced in a $10^\circ$ range, at  $\theta\simeq 40 \pm 5^\circ$.  This range is set both by the initial aspect ratio of the torus, which we take to be $\sim1$, and by the thickness of the X-ray heated 'skin' of the torus.  More optically thin models produce warm absorber-like spectra for $\theta\simeq 40 \pm 10^\circ$ , as they potentially provide more partially optically thin gas for evaporation. 

The bulk of the gas in this scenario has a terminal velocity of the order of the escape velocity at the inner torus edge.
Because of the funnel mechanism part of the gas is re-distributed to lower inclinations and acquires a higher terminal speed, $\sim 1000 \,\rm km\, s^{-1}$. In a real AGN environment such flow may contain clumps and irregularities and even dust, which are not captured in our studies because of the intrinsic limitations our methods. 
Accounting for the multiple phases of a gas (on a subcellular level) may reveal this in more detail and may also broaden the range of angles where the warm absorbers appear. 

The part of the flow that is shielded by the optically thick part of the torus body can also flow out as part of a torus global expansion.  Thus it strongly depends on the deposition of energy directly to its interior.
This problem is related to one of the infrared support of the AGN torus vertical structure against gravitational collapse \citep{Krolik07} and also requires additional investigation.

\hbox{}
This research was supported by an appointment to the NASA Postdoctoral Program at the NASA Goddard Space Flight Center, administered by Oak Ridge Associated Universities through a contract with NASA, and by grants from the NASA Astrophysics Theory Program 05-ATP05-18. We would like to thank the referee for his/her many 
constructive comments, which have lead to improvement of the manuscript.


\end{document}